\documentclass[aps,prd,twocolumn,superscriptaddress,nofootinbib]{revtex4}
\usepackage[pass,letterpaper]{geometry}
\pdfoutput=1  
\usepackage{graphicx,color}
\usepackage{latexsym,amsmath,amsfonts,amssymb,booktabs}
\usepackage[pdfencoding=auto]{hyperref}
\usepackage{slashed,upgreek,amscd,cancel,tensor,color}
\usepackage{url}
\usepackage{placeins}
\usepackage{doi}
\usepackage[normalem]{ulem}

\definecolor{MyBlue}{rgb}{0.15,0.15,0.70}
\hypersetup{
colorlinks=true,
citecolor=MyBlue,
linkcolor=MyBlue,
urlcolor=MyBlue
}

\newcommand{\be}{\begin{equation}}
\newcommand{\ee}{\end{equation}}
\newcommand{\beq}{\begin{equation}}
\newcommand{\eeq}{\end{equation}}
\newcommand{\bea}{\begin{eqnarray}}
\newcommand{\eea}{\end{eqnarray}}

\usepackage[stable]{footmisc}

\newcommand\ees{\end{eqnarray}}
\newcommand\bees{\begin{eqnarray}}

\begin{document}
\title{Comment on ``Velocity and Speed Correlations in Hamiltonian Flocks'' \\  by M. Casiulis {\it et al.} [{\tt arXiv:1911.06042}] }

\author{Andrea Cavagna}
\email{andrea.cavagna@roma1.infn.it}
\affiliation{Istituto Sistemi Complessi, Consiglio Nazionale delle Ricerche, Via dei Taurini 19, 00185 Rome, Italy}
\author{Irene Giardina}
\email{irene.giardina@roma1.infn.it}
\affiliation{Dipartimento di Fisica, Universit\` a Sapienza, P.le Aldo Moro 5, 00185 Roma, Italy}
\author{Massimiliano Viale}
\email{massimiliano.viale@cnr.it}
\affiliation{Istituto Sistemi Complessi, Consiglio Nazionale delle Ricerche, Via dei Taurini 19, 00185 Rome, Italy}

\date{\today}

\begin{abstract}
In {\tt arXiv:1911.06042v1} M. Casiulis {\it et al.} study a Hamiltonian model in which rigid rotations of moving clusters give rise to scale-free correlations of velocity and speed. M. Casiulis {\it et al.} compare correlations in their model to those observed in real flocks of birds and claim that rigid-body rotations provide an explanation that stands in contrast with, and it is simpler than, previously proposed explanations of correlations in bird flocks, namely Goldstone modes in the velocity orientations and marginal (or near-critical) modes in the speed. Here, we show that the rigid rotation scenario is completely inconsistent with a large body of well-established experimental evidence on real flocks of birds, and it therefore does not provide an appropriate explanation for the observed phenomenology.
\end{abstract}
\maketitle

In \cite{casiulis2019velocity} M. Casiulis {\it et al.} study a $2d$ Hamiltonian model that, at low temperature and intermediate densities, exhibits phase coexistence between a collectively moving droplet and a still gas. The conservation of angular momentum leads to rigid rotations of the droplet, which in turn produce velocity correlation functions that have a scale-free form (range of the correlation scaling with system's size $L$). Scale-free velocity correlations have also been observed in real flocks of birds \cite{cavagna+al_10}. M. Casiulis {\it et al.} analyze one flock in the dataset of \cite{cavagna+al_10} and claim that the data support the rigid-body rotation scenario also in real flocks, hence concluding that previous explanations of scale-free correlations in flocks, namely Goldstone modes in the velocity orientations \cite{cavagna+al_10,bialek+al_12,cavagna2018physics} and a marginal (or near-critical) mode in the speed \cite{bialek2014social, cavagna2019low}, are unnecessary. 

Here, we comment on these claims, showing that the rigid rotations described in \cite{casiulis2019velocity} are  inconsistent with all available experimental evidence. We also show that the visualisation method used by M. Casiulis {\it et al.} to present the experimental data of \cite{cavagna+al_10} erroneously conveys the visual impression of rigid-body rotations also in synthetic displacement fields that by construction are {\it not} generated by rigid-body rotations. Finally we present experimental data showing that, even once the rigid rotation is subtracted, the scale-free nature of correlations in flocks remains exactly the same.


\subsection{Real flocks do not turn by the parallel path rotations of \cite{casiulis2019velocity}}

The rotations displayed by the model of M. Casiulis {\it et al.} \cite{casiulis2019velocity} are rigid {\it parallel-path} rotations, which are just the standard rotations in classical mechanics \cite{goldstein}. In a rigid parallel path rotation, the trajectories of the points are all parallel to each other, they do not cross; the radii of curvature of the particles are given by the actual distances of the particles from the unique centre of rotation, hence they are different from each other. According then to the obvious equation, $v=\omega r$, the speed of the particles in a rigid parallel path rotation is larger at the external side of the turn, and smaller at the internal side, as it indeed happens in the model of \cite{casiulis2019velocity}. Note that it is impossible to operate a rigid parallel-path rotation of a set of points in which all points have the same speed.

There is a different type of rotation, called {\it equal-radius} rotation, in which each particle follows a path with the same radius of curvature, around a centre of rotation whose poistion is specific to each particle, so that individual paths cross at different times \cite{cavagna+al_15}. It is important to note that equal-radius is the only way to turn when each particle has {\it fixed speed}: to keep cohesion the overall angular velocity, $\omega$, must be the same, hence the equation $v=\omega r$ prescribes that all radii of curvature must be the same if all speeds are the same. The reader can watch Video3.mp4 included in this submission for a visualisation of an equal-radius turn at fixed speed (from \cite{cavagna+al_15}).\footnote{From a mathematical point of view, parallel path (or rigid body) rotations are generated by standard (orbital) angular momentum $L$, which rotates positions around the origin of the reference frame, while equal radius rotations are generated by spin $S$, which rotates the order parameter (i.e. the velocity in the case of flocks), rather than the position, of each bird \cite{benedetto2019some}. The angular momentum generates rotation in the external space of positions (or world sheet), while the spin generates rotations in the internal space of the order parameter \cite{attanasi+al_14}. This difference has been exploited to introduce a novel model of flocking that explains and fits very well the propagation of turns in real flocks \cite{cavagna+al_15}.}

There is by now ample and very compelling experimental evidence that real flocks turn collectively through equal-radius rotations, and not through rigid parallel path rotations \cite{pomeroy1992structure, ballerini+al_08b, attanasi+al_14, attanasi2015emergence, ling2019collective}. Even though, of course, animals do not have strictly fixed speed, biological groups turn by equal radius, and not by rigid parallel paths, for a very good reason: the speed of an animal can only fluctuate around some species-specific value, fixed by physiological and environmental constraints. Hence, a way of turning that requires a speed proportional to the turning radius would either force the external animals to move at unreasonably high speeds, or it would cut-off the size $L$ of the group severely, so to limit the radii of curvature to be compatible with the finite biological fluctuations of the speed. Both scenarios are biologically very unrealistic. Therefore, the physical backbone of the model of M. Casiulis {\it et al.} contrasts with the experimental evidence and with the biological reality of natural flocks.

\subsection{Real flocks do not have the ordered configurational structure of the clusters in \cite{casiulis2019velocity}}

The rigid-body structure emerging in the model of \cite{casiulis2019velocity} is due to the presence of a short-range repulsive potential, and also to an effective ferromagnetic-induced attraction. This potential is metric, namely it contains an intrinsic scale of length $r_c$, hence producing, at low temperature, solid-like crystalline structures, with quite large hexatic order parameter \cite{casiulis2019velocity}. Even though M. Casiulis {\it et al.} do not report configurational correlations, such a solid-like structure is certainly characterized by a very peaked and structured radial correlation function, $g(r)$ \cite{hansen_book}. Experiments, however, demonstrate that flocks have a very bland form of $g(r)$, which is far less structured than that of a liquid, not to mention that of the crystal configurations of \cite{casiulis2019velocity}; the $g(r)$ in real flocks actually resembles that of a gas \cite{cavagna+al_08c}.\footnote{This experimental result is probably in line with the fact that several studies \cite{ballerini+al_08, bialek+al_12, ling2019collective} show that flocks are ruled by a topological, and not by a metric, interaction: given a focal bird, interaction in flocks decays as a function of the number of neighbours, not of their actual metric distance.} Hence, the configurational order at the basis of the rigidity of model \cite{casiulis2019velocity} is completely incompatible with experimental data in natural flocks.

\subsection{The slow network rearrangement in real flocks does not imply parallel-path rotations of \cite{casiulis2019velocity}}

According to M. Casiulis {\it et al.}, the first experimental observation backing the rigid parallel-path rotation scenario is the following \cite{casiulis2019velocity}: ``On short time scales, the flocks do not rearrange: they are solid". Here M. Casiulis {\it et al.} simply write ``solid", but it is important to remark again that the solid-body rotations of model \cite{casiulis2019velocity} are a particular sub-class of solid-like rotations, namely, as we have already remarked, rigid parallel-path rotations. Hence, in order for the quoted experimental evidence to support the model of M. Casiulis {\it et al.}, one should rephrase it more precisely as: ``On short time scales, the flocks do not rearrange: they perform rigid parallel-path rotations". Let us see whether this is correct. Truly enough it was found in \cite{mora2016local} that the time needed to rearrange significantly the local interaction network in real flocks is much larger than the time needed to relax the local order parameter. This means that even during a turn the neighbours of one given bird remain approximately the same, so that the static ferromagnetic interaction (alignment) does not change significantly. However, this is {\it not} the same as saying that the system performs a rigid parallel-path rotation. When the system undergoes an equal-radius rotation, the neighbours of a given particle remain overall the same, but their {\it mutual orientations change}, in a fashion that is impossible to achieve in a parallel-path rotation, which conserves the mutual orientation of the particles. Such change in the mutual orientation of the neighbours during turns has been observed experimentally in a very clear way \cite{attanasi2015emergence}. Therefore, the identification made in \cite{casiulis2019velocity}, namely that no change of neighbours during a turn implies solid-body parallel-path rotation, is incorrect.

\begin{figure}[!htb]
\includegraphics[width=121pt]{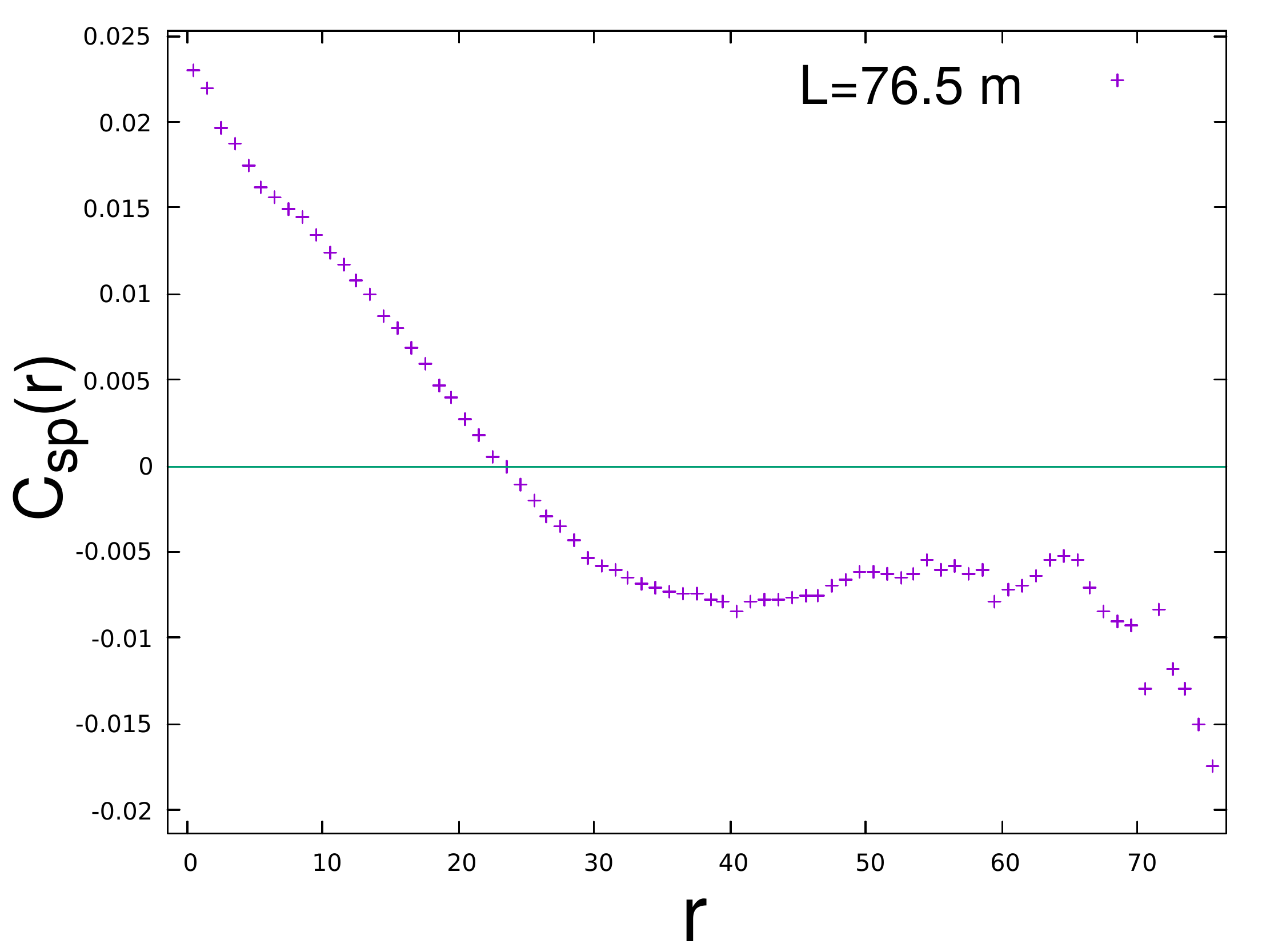}
\includegraphics[width=121pt]{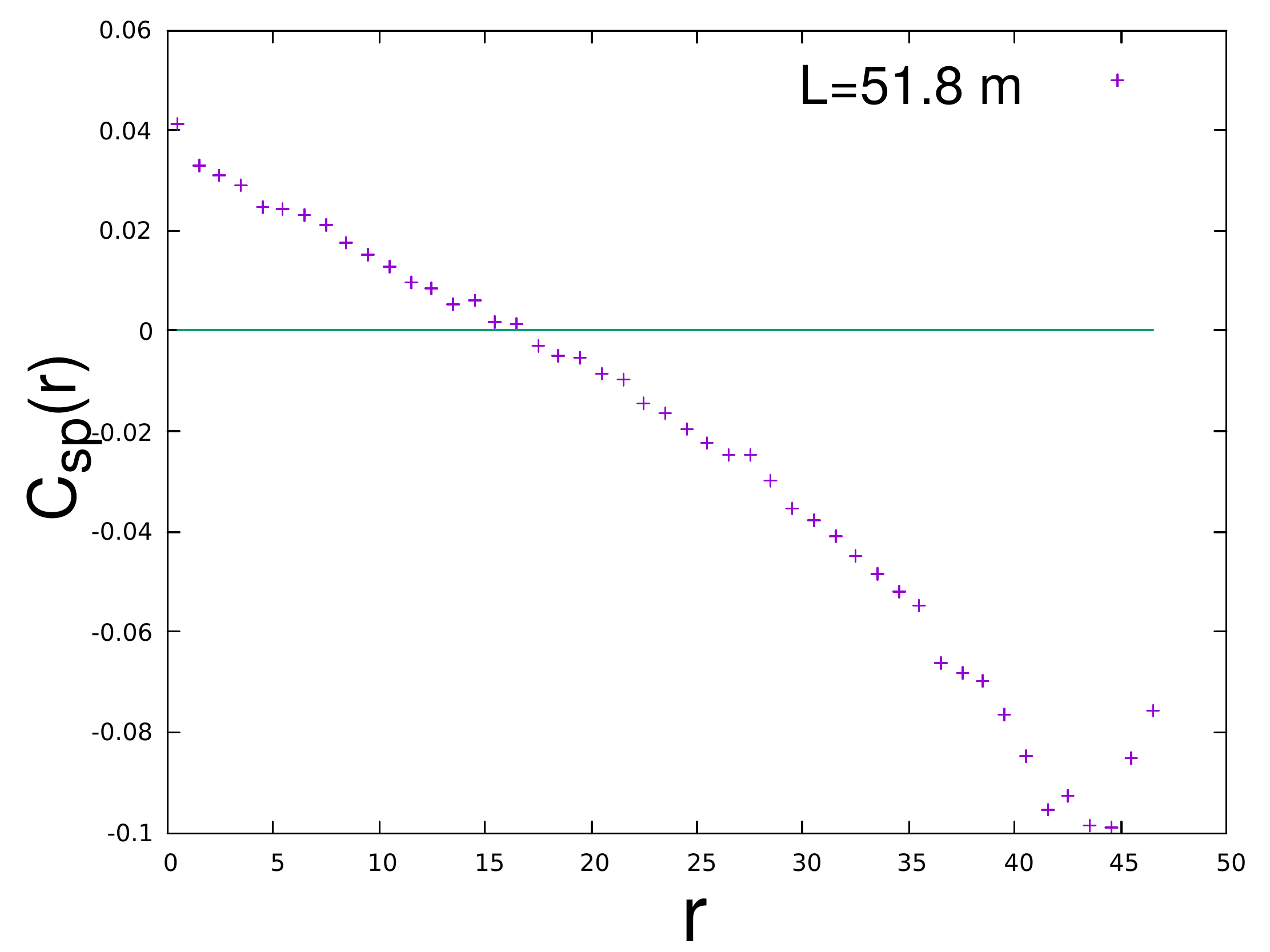}\\
\includegraphics[width=121pt]{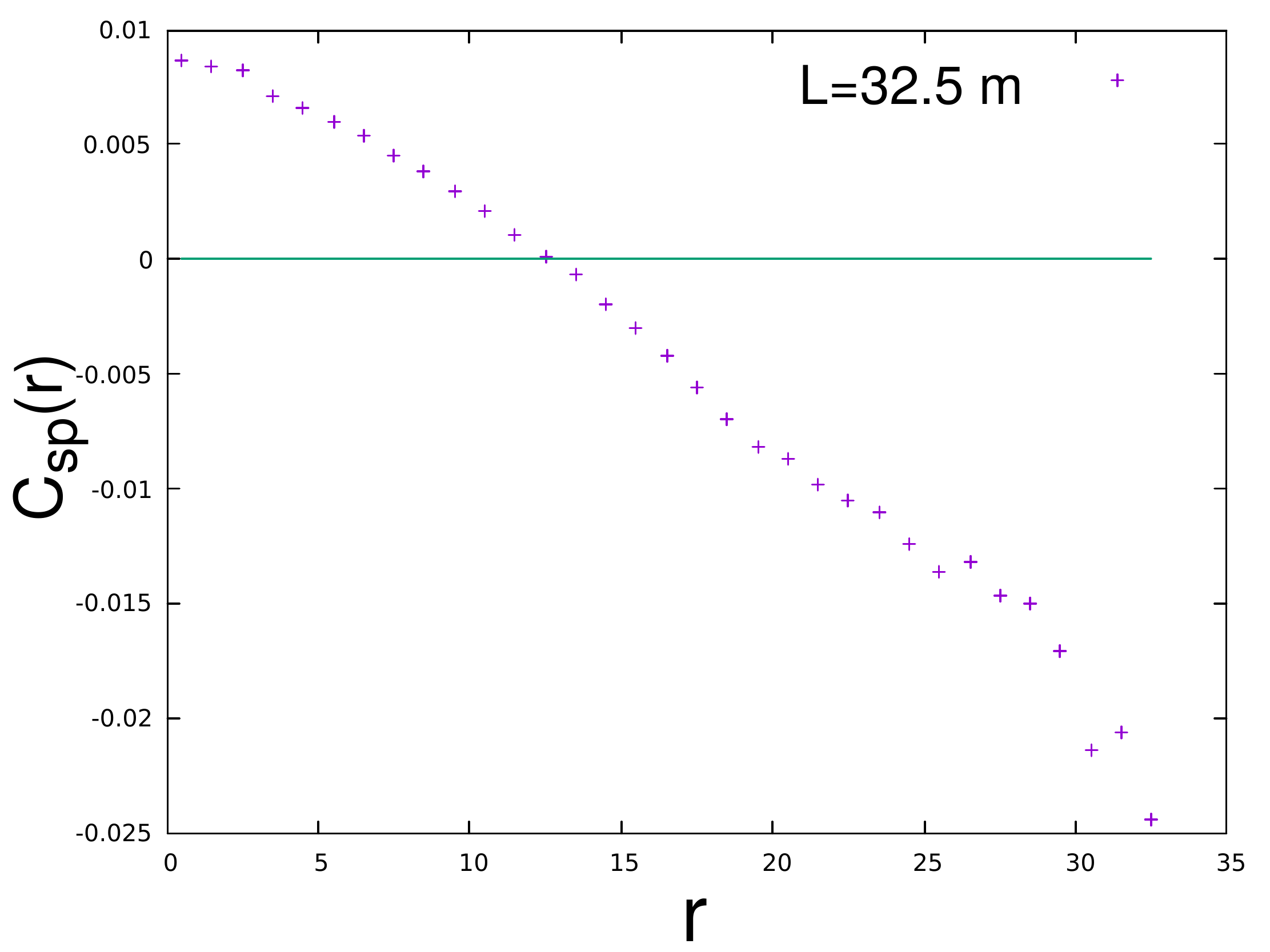}\includegraphics[width=121pt]{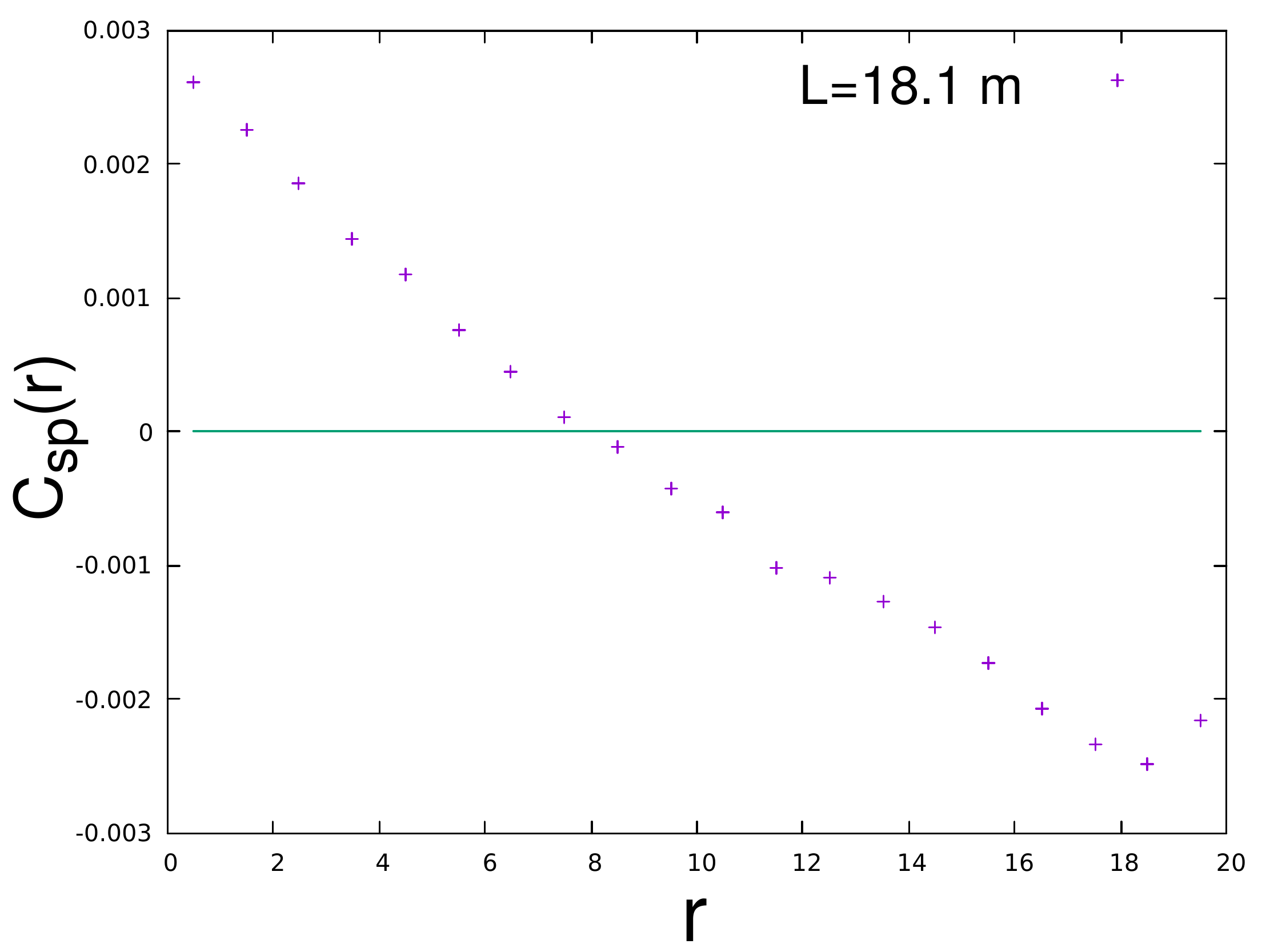}
\caption{Experimental speed connected correlation function, in flocks 31-01, 17-06, 21-06, 58-07, from the database of \cite{cavagna+al_10}. The shape of the correlation is completely different from the rigid-rotating disk one described in \cite{casiulis2019velocity}, in which the function climbs up for large $r$ (see Fig. SI3b of \cite{casiulis2019velocity}). The flocks database of \cite{cavagna+al_10} contains very many events of this type.}
\label{fig:bio-speed}
\end{figure}

\subsection{Experimental speed correlations in real flocks do not typically have the rigid-disk shape of \cite{casiulis2019velocity}}

The second experimental observation supposedly backing the rigid-body rotation scenario according to M. Casiulis {\it et al.} is that the rigid disk rotation studied analytically in \cite{casiulis2019velocity} produces  correlation function of the speed fluctuations, $C_\mathrm{sp}(r)$, that rise up for large $r$, which is also what happens in the experimental acquisition 28-10 of \cite{cavagna+al_10}. However, this is not a generic behaviour of real flocks: in acquisition 25-10 reported in Fig.2 of \cite{bialek2014social} (another large flock of more than $1000$ birds), for example, one observes a completely different shape of the speed correlation function. In fact, the same thing happens in several other flocks in our database (see Fig.\ref{fig:bio-speed}). Therefore, the rigid-disk rotation does not fit the phenomenology of real flocks. In reality, the shape of the correlation function for large $r$, far from being a general unifying trait, is dominated by very specific features of the flock under considerations, as the border shape, the main axis vs the direction of motion, and the phase gradient axis.

\subsection{The visualisation of \cite{casiulis2019velocity} conveys the impression of parallel-path rotation also when there is none}

\begin{figure}[!htb]
\includegraphics[width=160pt]{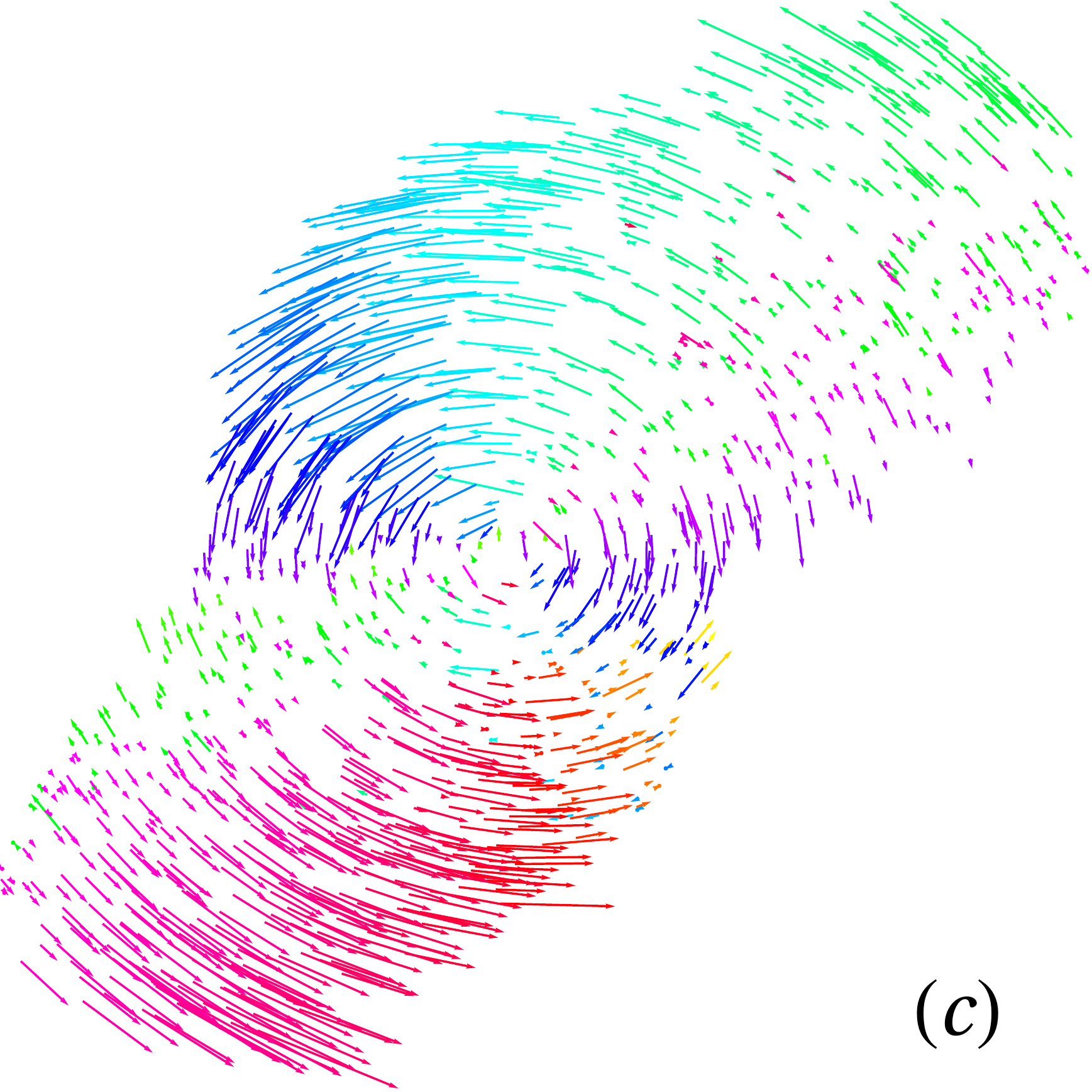}
\caption{Tangential velocity fluctuations (called ``azimutal'' in \cite{casiulis2019velocity}), for a real flock of birds (figure from \cite{casiulis2019velocity}; each vector is multiplied by an (arbitrary) rescaling factor ``such that all displacements are visible''.)}
\label{fig:loro}
\end{figure}

The third piece of experimental evidence supposedly supporting the rigid parallel-path rotation scenario of \cite{casiulis2019velocity} comes from a visualisation method of the displacement field applied by M. Casiulis {\it et al.} to flock 28-10 in the database of \cite{cavagna+al_10}. More specifically, in Fig.SI3 of \cite{casiulis2019velocity} (reproduced in Fig.\ref{fig:loro} here), some ``bird displacement fluctuations'' are presented, which strongly convey the impression that parallel-path rigid-body rotations are indeed present in this real flock. M. Casiulis {\it et al.} write: ``Although it is not a perfect solid rotation, this picture shows that there is indeed some rigid rotation in the flock, which could account for the correlations akin to those observed in our system.''  Given our discussion of equal-radius turns in flocks, this result seems surprising: if flocks turn in equal radius fashion, how is it possible that the displacement field looks so similar to the parallel-path displacement field of a  rotating disk? We explain below that this result is an artefact of the method used in \cite{casiulis2019velocity} to define  the displacement fluctuations.

To visualise the parallel-path rotation present in a displacement field M. Casiulis {\it et al.} use the following procedure: it is first computed the global angular momentum in the reference frame of the center of mass and at the projections of velocity fluctuations and positions on the plane orthogonal to its direction. At this point one is dealing with two-dimensional vectors. M. Casiulis {\it et al.} then define the tangential components $\bf \delta v_\mathrm{tan}$ of the velocity fluctuations (what M. Casiulis {\it et al.} call the azimuthal components); more precisely, 
\beq
\delta{\bf v}_\mathrm{tan} = \frac{|\delta {\bf v} \times \delta{\bf r}|}{|\delta{\bf r}|} {\bf n_\mathrm{tan}}
\eeq
where $\delta {\bf v}$ and $\delta{\bf r}$ are the velocities and positions in the centre of mass reference frame (projected on the plane), and ${\bf n_\mathrm{tan}}=(-\delta y/\delta r, \delta x/\delta r)$ is the tangential unit vector. Second, each one of these vectors is multiplied by an arbitrary rescaling factor ``such that all displacements are visible'' \cite{casiulis2019velocity} and plotted as a vector field. The result is the field in Fig.\ref{fig:loro}, which admittedly looks remarkably similar to a rigid disk displacement field. Let us now see how this same method works in a synthetic case, completely under analytic control.

Experiments show that, when a turn occurs, the perturbation in the phase of the velocity starts at some localized position in the flock, and then it propagates to the rest of the system \cite{attanasi+al_14}. This means that at any given time during a turn there is a phase gradient crossing the flock. Because a phase gradient means a certain degree of local misalignment between the birds velocities, the wave length of such a phase gradient is typically large, of the same order as the system's size \cite{cavagna2018physics}.\footnote{
This is, of course, the obvious mechanisms behind classical spin waves in systems with spontaneously broken continuous symmetry, and therefore behind the Goldstone theorem \cite{halperin1969hydrodynamic}. Yet we do not need to invoke such a sophisticated piece of theoretical physics here, but merely to remark that flying $\pi$ off your nearest neighbours is not a good idea in a packed and fast group.}

We now build the most elementary example based on the above observation. We consider two set of points in $2d$, with the following ingredients: {\it i)} all displacements have exactly the same modulus, as if all particles had the same speed; {\it ii)} a smooth phase gradient crosses the system, with a wavelength of the same order as the system's size.  Point {\it i)} is the most crucial: we know that in real flocks there are speed fluctuations, yet here we want to build two configurations of points that {\it by construction} are not generated by a rigid parallel-path rotation; with fixed speed there is no way a rigid parallel-path rotation can be implemented. As a consequence, it is crucial to notice that the speed correlation function {\it is exactly zero} in this example, as there are no speed fluctuations what-so-ever.

\begin{figure}[!htb]
\includegraphics[width=270pt]{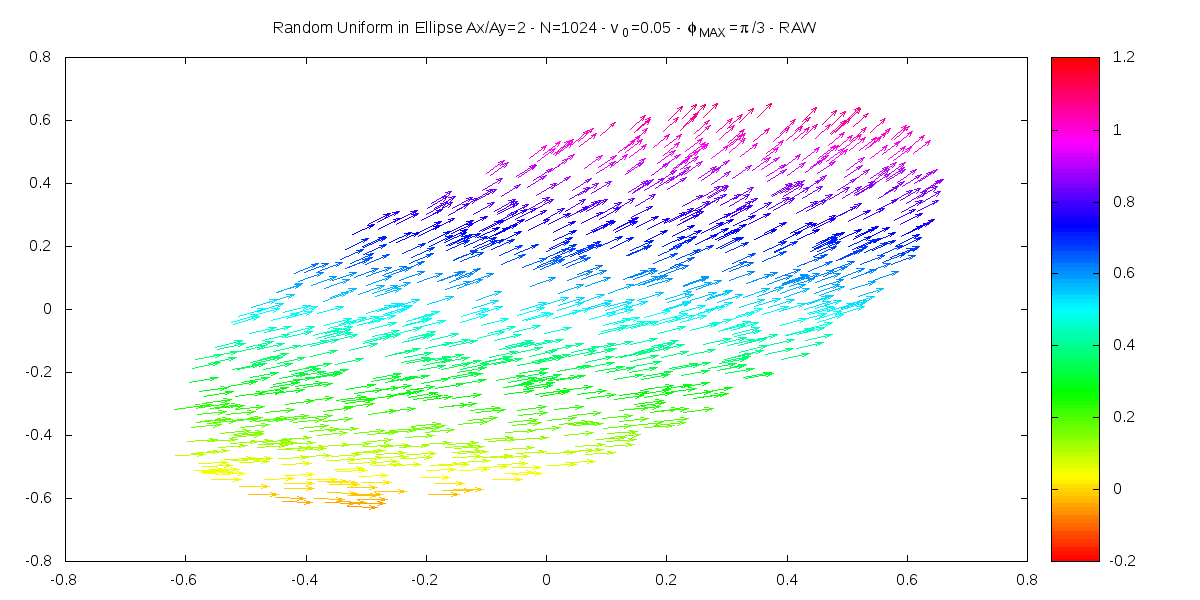}
\caption{Synthetic velocity field in $2d$, with constant modulus and a phase gradient along the $y$ axis (eq.\eqref{field}).}
  \label{fig:ellipse}
\end{figure}

The following displacement field, represented in Fig.\ref{fig:ellipse}, serves to our purpose; (with a slight abuse of notation, we use the letter $v$, even though these are displacements),
\bea
v_x(x,y) &=& v_0 \cos(y\pi /3L)
\nonumber
\\
v_y(x,y) &=& v_0 \sin(y\pi/3L)
 \, ,
\label{field}
\eea
with speed (modulus of displacement) $v_0=0.05$, for all points. `Birds' in our example are randomly distributed within an ellipse, just to make the example more flockish-looking, but this is completely irrelevant of course. The phase gradient along the $y$ direction is such that the total phase change from one side to the other of the system is $\pi/3$. 

\begin{figure}[!htb]
\includegraphics[width=270pt]{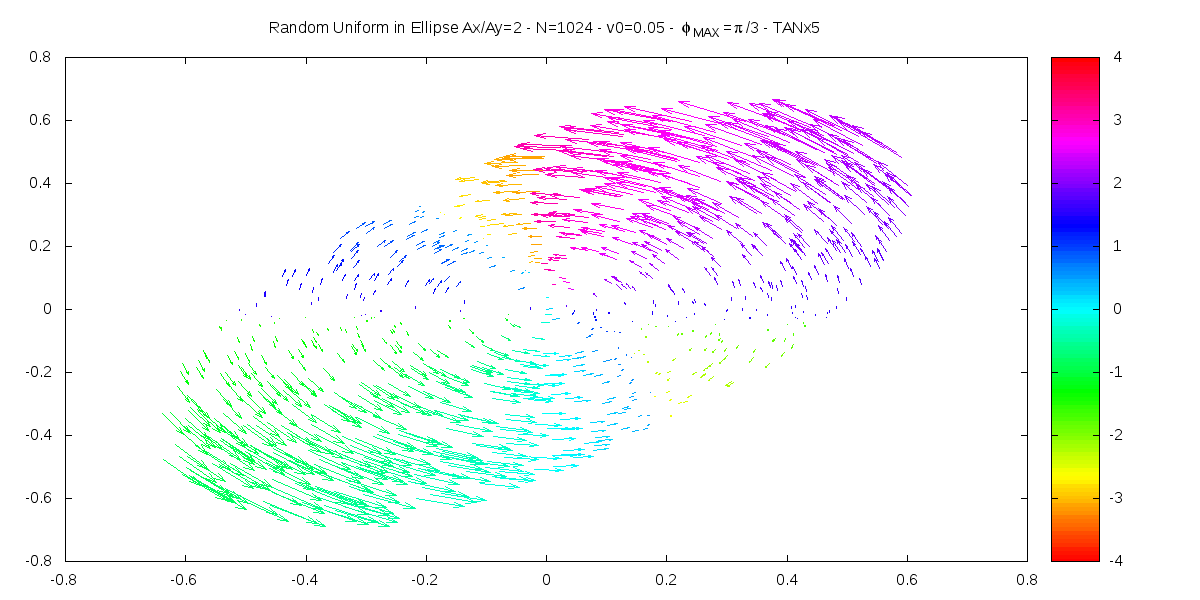}
\caption{Tangential velocity fluctuations (called ``azimutal'' in \cite{casiulis2019velocity}) as computed in \cite{casiulis2019velocity}, for the field in Fig.\ref{fig:ellipse}; as in \cite{casiulis2019velocity} each vector is multiplied by an (arbitrary) rescaling factor ``such that all displacements are visible''.}
  \label{fig:tangential}
\end{figure}

Let us see what happens if we apply the method of M. Casiulis {\it et al.} to our synthetic field, eq.\eqref{field}-Fig.\ref{fig:ellipse}.
The result of this procedure can be seen in Fig.\ref{fig:tangential}. This displacement field looks {\it exactly} like the one presented in Figs.SI-3c of \cite{casiulis2019velocity} (Fig.\ref{fig:loro} here), conveying the strong impression that this pattern comes from an underlying parallel-path rigid-disk rotation, while it does not. We stress again that the ones we consider in \eqref{field} are two configurations of points which by construction are {\it not} connected by a parallel path rigid-body rotation, as all displacements have the same modulus. It is easy to check that exactly the same similarity occurs for the radial displacement field of \cite{casiulis2019velocity}. We conclude that the visualisation tool employed in \cite{casiulis2019velocity} is quite faulty, as it conveys the strong impression of an underlying rigid parallel-path rotation even in cases where there is none.

\subsection{Experimental correlations in real-flocks are scale-free even after subtracting the fitted parallel-path rotation}

If one is not convinced by any of the arguments above, one could still ask what happens if one subtracts from the experimental data the optimal rigid rotation and computes the velocity correlation on this new data-set. The standard procedure to fit rotations to two sets of points is Kabsch's method  \cite{kabsch1976solution}, which in $2d$ is particularly elementary: it simply finds the angle $\theta$ that minimizes the RMSD between the first set of points and the rotated second set of points. We emphasise once again that the rigid rotation fitted by Kabsch method is more precisely a parallel-path rigid rotation, namely a classic one-matrix rotation around a single centre \cite{kabsch1976solution}. Notice also that Kabsh method first finds the optimal translation connecting the two sets of points, and then it fits the optimal rotation in the centre-of-mass reference frame \cite{kabsch1976solution}.

\begin{figure}[!htb]
\includegraphics[width=250pt]{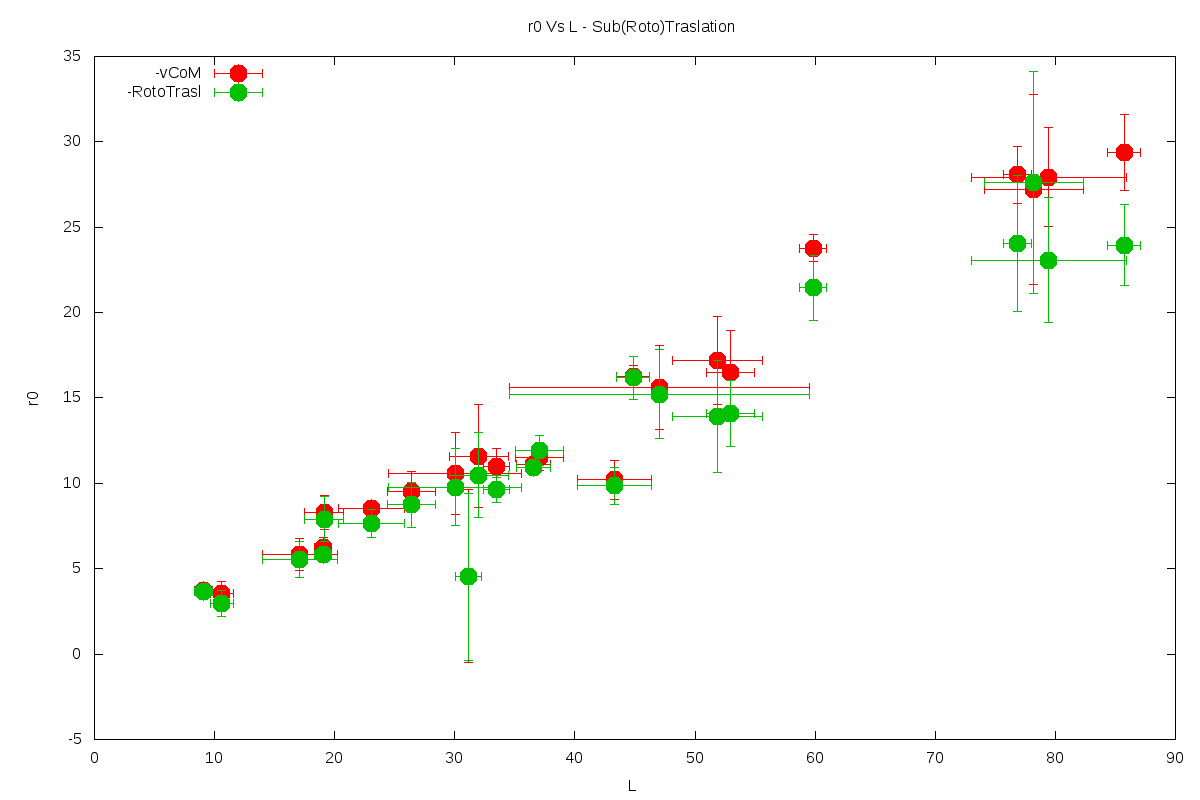}
\caption{Correlation length vs. flock's size in experimental data. Each point corresponds to a flocking event and it is an average over several instants of time in that event, error bars corresponds to SDs. Red: the velocity fluctuations are calculated as in \cite{cavagna+al_10} by just subtracting the optimal translation (i.e. the centre of mass displacement). Green: the velocity fluctuations are calculated by subtracting both the optimal translation and the optimal rigid rotation (Kabsch algorithm). The scale-free nature of the correlation is the same in the two cases. Original data from \cite{cavagna+al_10}. }
\label{fig:scale-free}
\end{figure}

We therefore re-analyzed the data in \cite{cavagna+al_10} by subtracting the Kabsch rigid rotation from the displacement field and then recalculating the connected correlation functions  for each flocking event.  After doing this we find no significant difference with respect to the original published data in the scaling with $L$ of the correlation length (see Fig.\ref{fig:scale-free}). Even though we do not report them here, the individual shape of the correlation functions after subtracting the rotation is very much the same as before.\footnote{Notice that, due to the equal radius nature of flock's turning, the global rotation and translation fitted by Kabsch is always very small, which is also the case for the field \eqref{field}. Hence, it is not surprising that the correlations do not change much.} 
We believe that this last result proves, in quite neutral an empirical way, that rigid parallel-path rotations are irrelevant in describing the origin of velocity correlations in bird flocks.

\subsection{Conclusions}

The solid-body rotations present in the model of M. Casiulis {\it et al.} \cite{casiulis2019velocity} are rigid parallel-path rotations, in which each point is rotated around a single centre, with radii of rotations, and therefore speeds, that vary from point to point. We have presented ample evidence that parallel-path rotations are completely alien to real flocks phenomenology. Let us summarize such evidence here: {\it i)} real flocks are known to turn according to equal-radius rotations, and not to parallel-path rotations, for obvious physiological reasons; {\it ii)} real flocks show none of the strong configurational order in the $g(r)$ responsible for the rigid-body rotation of \cite{casiulis2019velocity}; {\it iii)} the rearrangement of the mutual orientations of the neigbours during turns in real flocks is totally incompatible with the parallel-path rotations of \cite{casiulis2019velocity}; {\it iv)} in very many real flocks, the large $r$ shape of the speed correlation function is different from the rigid disk one presented in \cite{casiulis2019velocity}; {\it v)} the visualisation tool of M. Casiulis {\it et al.} conveys the impression of rigid parallel-path rotations even when there is none; {\it vi)} even if one insists in subtracting a rigid rotation from the displacement fields of real flocks, scale free correlations do not change at all. We conclude that the rigid rotation framework presented by M. Casiulis {\it et al.} in \cite{casiulis2019velocity} is not the most compelling way of explaining scale-free correlations in natural flocks of birds.


\bibliography{general_cobbs_bibliography_file}

\end{document}